\documentclass[a4paper,11pt]{article}
\usepackage{jheppub}
\usepackage{lineno}

\usepackage{graphicx,epsfig}
\usepackage{amsmath}
\usepackage{amsfonts}
\usepackage{fancyhdr}
\usepackage{siunitx}
\usepackage{tikz}
\usepackage{appendix}
\usepackage{hyperref}
\usepackage[compat=1.1.0]{tikz-feynman}
\tikzfeynmanset{warn luatex=false}
\usepackage{comment}

\title{Universality of entanglement in gluon dynamics}

\author[a]{Claudia Núñez}
\affiliation[a]{Institut Cartogràfic i Geològic de Catalunya, Passeig de Santa Madrona, 45, 08038 Barcelona, Spain}
           
\author[b]{Alba Cervera-Lierta}
\affiliation[b]{Barcelona Supercomputing Center, Plaça Eusebi G\"uell, 1-3, 08034 Barcelona, Spain}            
\author[c]{José Ignacio Latorre}
\affiliation[c]{Centre for Quantum Technologies, National University of Singapore, Singapore}

\abstract{Entanglement of fundamental degrees of freedom in particle physics is generated {\sl ab initio} in scattering processes. We find that in the case of a pure $SU(N)$ gauge theory, two gluons in a product state can become maximally entangled in their  polarizations as the result of three- and four-gluon vertex interactions. 
Remarkably, the amount of entanglement among gluon polarizations is independent of the color degree of freedom. We also find that a small deviation of the relative weight between three- and four-gluon vertices would prevent the generation of maximal entanglement. This can be seen as a small piece of a possible {\sl it from qubit} principle  underlying fundamental interactions.
}

\begin{document}
\maketitle
\flushbottom

\section{Introduction}\label{sec:intro}

Entanglement is a technical word reserved to describe correlations in quantum mechanics, specifically those emerging for a quantum  state made of two or more subsystems that cannot be described as a classical combination of the states of each subpart.  Entanglement is a core feature in quantum physics that, by means of violation of Bell  inequalities \cite{Bell1964on},  discriminates between classical and quantum physics.

A question arises on the origin of entanglement in the basic processes of Nature. Can the fundamental interactions in the Standard Model create entangled states?  If so, do they provide mechanisms to obtain maximal entanglement from non-entangled states? The ultimate question would correspond to understand whether entanglement may turn to be a candidate to formulate a novel principle in physics, one demanding that physics is quantum, not classical.

This is the thread of thought explored in Refs. \cite{CerveraLierta2017maximal,CerveraLierta2019quantum} and the present work. The authors focused in QED and weak interactions, studying the correlations between helicity states in two-body scattering processes and decays.  For QED interactions, it was found that maximal entangled states are created from a product state by two mechanisms: $s$-channel processes at high energies where the virtual photon carries equal overlaps of the helicities of the final state particles; and the indistinguishable superposition of $t$ and $u$-channels, valid for all energies. The latter mechanism justifies why the low-energy interaction between two spins, namely the Heisenberg model, is able to generate maximal entanglement. It was also shown that requiring the generation of maximal entangled states leads to reproducing the exact QED photon-electron vertex. Such a result suggests the idea of exploring some kind of Maximal Entanglement Principle (MaxEnt) as a guiding element to  construct quantum theories. Finally, it was observed that maximal entanglement favors a  weak mixing angle of $\frac{\pi}{6}$, very close to the Standard Model value. A similar result is obtained in  Ref. \cite{Morales2024tripartite}, where maximal entanglement also favors a weak mixing angle of $\frac{\pi}{6}$ for the three-body Higgs boson decay $H \rightarrow \gamma l \bar{l} $ at 1-loop level. The generation of entanglement in QED scattering processes has also been studied in Refs.\cite {Fedida2023tree-level, Blasone2024entanglement, Blasone2024complete} where the authors do not restrict to initial product states. 

Further work has also been done in studying entanglement in positronium \cite{Acin2001three-party}, charmonium \cite{Baranov2008bells,Chen2013testing} and Higgs boson \cite{Barr2022testing,AguilarSaavedra2023laboratory-frame,AguilarSaavedra2023testing,Fabbrichesi2023bell} decays, generation of kaon \cite{Bramon2002novel}, B meson \cite{Takubo2021feasibility}, $\tau$ lepton \cite{Ehataht2024probing} and top quark \cite{Severi2022quantum,Fabbrichesi2021testing, Afik2022quantum} pairs, as well as in neutrino oscillations \cite{Blasone2009entanglement,Banerjee2015a} and vector boson scattering \cite{Morales2023exploring}, to propose Bell tests that could be experimentally verified. For top quarks, there have also been studies in quantum tomography techniques \cite{Afik2021entanglement} and, recently, entanglement between a top quark pair has been experimentally detected by the ATLAS and CMS collaborations at the LHC \cite{ATLASCollaboration2024observation,CMSCollaboration2024observation,CMSCollaboration2024measurements}. 
The production of these top quarks comes primarily from gluon interactions in these collisions. Then, since top quarks decay faster than their hadronization, they transfer their spin properties to their decay products, which allows the estimation of their entanglement properties from the measurement of the angular dependence of the detected jets. Therefore, the phenomenology surrounding gluon scattering is of special interest for high-energy physics experimentalists.
Violations of Bell inequalities have also been obtained experimentally in charmonium \cite{Fabbrichesi2024charmonium} and B meson \cite{Fabbrichesi2024bell,Gabrielli2024entanglement} decays. The use of entanglement in particle interactions has also been proposed to constrain new physics beyond the Standard Model using LHC measurements  \cite{Aoude2022quantum, Severi2023quantum, Aoude2023probing, Bernal2023entanglement, Fabbrichesi2023stringent, Maltoni2024quantum, Maltoni2024tops}.

We should also note a line of research complementary to the study of a possible MaxEnt principle, where the interplay between entanglement suppression in scattering processes and the emergence of global symmetries has been explored in Standard Model \cite{Beane2019entanglement, Low2021symmetry, Liu2023minimal} and beyond the Standard Model \cite{Carena2023entanglement, Kowalska2024entanglement}
interactions, as well as its relation with symmetry-breaking effects as quark and lepton mixing \cite{Thaler2025flavor}, and MaxEnt \cite{Kowalska2024entanglement}. 

In this work, we take a step further focusing on pure Yang-Mills gluon dynamics. We compute the polarized amplitudes for gluon scattering at tree-level using the three- and four-gluon couplings. Then, using the concurrence as a figure of merit, we show that entanglement is only generated when the initial product state presents opposite polarizations. Maximal entangled states are only produced in the case where the scattering angle is $\theta = \frac{\pi}{2}$. Then, the final states are always maximally entangled, independently of the color of the gluons involved in the process.  This result points at some  structure in pure Yang-Mills theory that imposes a sort of universal creation of entanglement, independent of the particular gauge group at play. 

It is tantalizing to investigate whether the relation between the three- and four-gluon vertices, as dictated by gauge symmetry, can be imposed from a MaxEnt Principle.
While this is not the case in full generality, we demonstrate that a clear and robust relationship does emerge.
This result suggests a deep relation between local symmetries and entanglement.

It must be made clear that gluons are not asymptotically free particles. It is thus not possible to perform a Bell test based on the gluon polarizations, as it is done for photons. The idea, though, is that maximal entanglement is indeed generated and conditions the subsequent evolution of the full system. In this sense, we here analyze the conditions for maximal entanglement to be generated.

The structure of this paper goes as follows. In Sect. \ref{sec:quant_ent},  we introduce a figure of merit to quantify entanglement for two particle scattering processes. In Sect. \ref{sec:tree-level} we present the results obtained for the total polarized scattering amplitudes in the gluon scattering. Sect. \ref{sec:generation_entanglement} is centered on the analysis of the generation of entanglement in these processes. Sect. \ref{sec:MaxEntk} is devoted to verify the way the relative weight between three- and four-gluon vertices affect entanglement, and shows how a MaxEnt principle works in this scenario. Our conclusions are presented in Sect. \ref{sec:conclusions}. Some additional information and conventions are included in App. \ref{app:conventions}. App. \ref{app:amplitudes} collects the complete set of polarized amplitudes computed for each channel. App. \ref{app:amplitudes4k} lists the complete set of polarized amplitudes when the balance between the 3- and 4-gluon vertices is modified. We should note that Ref. \cite{CerveraLierta2019quantum} also presents an analysis of the gluon scattering process. The present work presents new results and conclusions.

\section{A figure of merit for entanglement} \label{sec:quant_ent}

In order to quantify entanglement in gluon scattering  it is necessary to specify the quantum degrees of freedom at stake as well as to provide a precise figure of merit. We shall discuss entanglement in terms of polarizations of gluons, and entanglement will be quantified using the concurrence obtained from the coefficients of the superposition of final states. 

 Considering that the polarization of the gluons can take two values, right-handed ($R$) and left-handed ($L$), we can describe the incoming and outgoing states as two qubit states with basis $\{|R\rangle, |L\rangle\}$. After the interaction, the final state will be a superposition of all possible combinations of the two polarizations. Therefore, for an initial product state $|RR\rangle$, $|RL\rangle$, $|LR\rangle$ or $|LL\rangle$, the final state $|\psi_{f}\rangle$ can be written as
\small
\begin{eqnarray}
    |\psi_{f}\rangle &\sim& 
    \mathcal{M}_{\psi_{i} \rightarrow
    RR}|RR\rangle + \mathcal{M}_{\psi_{i}
    \rightarrow RL}|RL\rangle
     + \mathcal{M}_{\psi_{i}
    \rightarrow LR}|LR \rangle+
    \mathcal{M}_{\psi_{i} \rightarrow
    LL}|LL\rangle,
    \label{eq:finalstate}
\end{eqnarray}
\normalsize
where  $\mathcal{M}_{\psi_{i}\rightarrow AB}$ is the scattering amplitude at tree level for the process where the final state is $|AB\rangle$. Let us note that these amplitudes are a function of momenta, as well as the coupling constant. 

To quantify the entanglement of these states we use the concurrence as the  figure of merit. It is well-known that in two-level states, as is the case here, all ways of measuring entanglement reduce to a single combination. Given a two particle pure state
\begin{equation}
     |\psi\rangle = 
    \alpha|RR\rangle + \beta|RL\rangle + \gamma|LR\rangle +
    \delta|LL\rangle,
    \label{eq:psi_general}
\end{equation}
with $\alpha, \beta, \gamma, \delta \in \mathbb{C}$ and $|\alpha|^{2}+|\beta|^{2}+|\gamma|^{2}+|\delta|^{2} = 1$, the concurrence is defined as 
\begin{equation}
    \Delta = 2|\alpha\delta - \beta\gamma|,
    \label{eq:delta}
\end{equation}
where $0 \leq \Delta \leq 1$. The states with $\Delta = 0$ correspond to product states and the ones with $\Delta = 1$ to maximal entangled states. Thereby, we will start with initial states where the concurrence equals 0 and explore if there are any final states where this value is increased to 1. All the coefficients and, thus, the concurrence are a function of the coupling constant and the momenta defining the kinematics of the process.

\section{Tree-level gluon scattering amplitudes}\label{sec:tree-level}

\begin{figure} [ht]
\centering
\includegraphics[scale=0.8]{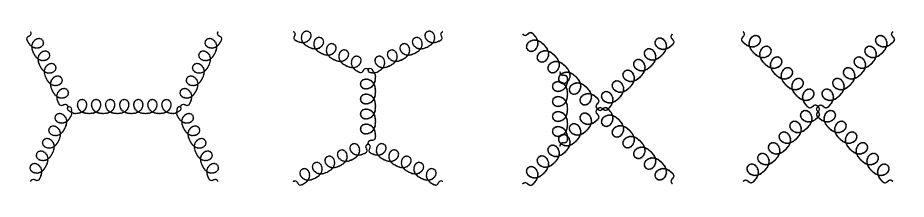}
\caption{Channels that contribute to the gluon scattering. From left to right, s-channel, t-channel, u-channel and quartic channel.}
\label{fig:chan}
\end{figure}

The gluon scattering process involves four Feynman diagrams that correspond to $s$, $t$, $u$ and 4-vertex channels, as shown in Fig.~\ref{fig:chan}.

We compute explicitly the amplitudes for the four channels, using the Feynman rules for these processes. A powerful method to obtain the total polarized amplitudes, that is the spinor helicity formalism \cite{Elvang2015scattering}, allows to compute scattering amplitudes considering only the external particles. It is then possible to get the final amplitudes in a straightforward way, avoiding long computations. However, we are interested in the relation between the different channels involved in the scattering process to analyze the detailed mechanisms that generate entanglement. For this reason we do not resort to the spinor helicity formalism. 

Let us now concentrate on the scattering amplitudes that are not null. The details of the kinematics, Feynman rules and conventions used are listed in App. \ref{app:conventions}. The complete set of polarized amplitudes computed for each channel are collected in App. \ref{app:amplitudes}. Using these values, the total amplitude is obtained by summing up the amplitudes for each channel
\begin{equation}
    \mathcal{M} = \mathcal{M}_{s} + \mathcal{M}_{t} + \mathcal{M}_{u} + \mathcal{M}_{4}.
    \label{eq:M_channels}
\end{equation}
All amplitudes carry a common prefactor involving the coupling constant that will cancel when computing the concurrence.

We shall consider two incoming gluons in a product state of polarizations, $|RR\rangle$, $|RL\rangle$, $|LR\rangle$ or $|LL\rangle$, and momenta $p_{1}$, $p_{2}$ and color $a$, $b$ respectively. After the interaction, we obtain a final state of the form Eq.~\eqref{eq:finalstate}, where the outgoing gluons are characterized by $p_{3}$, $p_{4}$ and $a'$, $b'$, respectively. 

If the initial state share the same polarization, the interaction does not change the polarization of the gluons and the final state remains the same product state, 
\begin{multline}
    \mathcal{M}_{RR \rightarrow RR} = 
        \mathcal{M}_{LL \rightarrow LL} \\= 
         2 g^{2} \Bigg[f^{abc}f^{a'b'c}\left(\frac{u-t}{s}\right) +
        f^{aa'c}f^{bb'c}\left(2 + \frac{u-t}{s}\right)\frac{u}{t} 
         + f^{ab'c}f^{ba'c}\left(2 - \frac{u-t}{s}\right)\frac{t}{u}\Bigg],\
         \label{eq:MRR}
\end{multline}
where $g$ is the strong coupling constant and $s$, $t$, $u$ the Mandelstam variables defined in Eq. (\ref{mandelstam}). The values $f^{abc}$ are the structure constants of the $SU(N)$ gauge theory, that are defined through the commutation relation between its generators $[T_{a}, T_{b}] = if^{abc}T_{c}$. In this scattering process, there is no generation of entanglement. 

When the initial product state have opposite polarizations, let it be $RL$ or $LR$, the interaction produces a superposition of polarization,
\begin{eqnarray}
    \mathcal{M}_{RL \rightarrow RL} &=&
        \mathcal{M}_{LR \rightarrow LR} =  -2 g^{2}\Bigg[f^{aa'c}f^{bb'c}\left(\frac{u^{2}}{ts}\right) + f^{ab'c}f^{ba'c} \left(\frac{u}{s}\right)\Bigg], \nonumber\\
     \mathcal{M}_{RL \rightarrow LR} &=&
        \mathcal{M}_{LR \rightarrow RL} = 
     -2 g^{2}\Bigg[f^{aa'c}f^{bb'c}\left(\frac{t}{s}\right) + f^{ab'c}f^{ba'c} \left(\frac{t^{2}}{su}\right)\Bigg].
    \label{eq:MRL}
\end{eqnarray}
To simplify the notation, we will use $F_1\equiv f^{aa'c}f^{bb'c}$ and $F_2\equiv f^{ab'c}f^{ba'c}$.

Using this shorthand notation, we now address the issue to write the
amplitudes as restricted to the subspace of two gluons. This can be
done by normalizing the state, so that the colour charge is dropped as well as global factors and signs
\begin{equation}
   |\psi\rangle  =  \frac{1}{\sqrt N}
   \left(
    \frac{u}{ts}F|RL\rangle + 
     \frac{t}{us}F|LR\rangle
    \right),
\end{equation}
where we use the shorthand notation $F=F_1 u + F_2 t$, and 
where  the normalization factor is
\begin{equation}
   N =  
   \left( \frac{u}{ts} F\right)^2  
   +\left( \frac{t}{us} F\right)^2
   =  F^2
   \frac{u^ 4+t^4}{s^ 2 t^ 2 u^ 2}.
\end{equation}
This normalization makes only sense if the amplitudes are not null due to their color indices. 

In the case of an initial $RL$ state, the above final amplitudes allow us to cast the effective final state in the subspace of polarizations in first-order perturbation theory
\begin{equation}
   |\psi\rangle_{RL \rightarrow RL + LR} =  \frac{1}{\sqrt{t^{4} + u^{4}}}\left( u^{2}|RL\rangle + t^{2}|LR\rangle\right)
   \label{eq:finalRL} .
\end{equation}
In the case of an initial $|LR\rangle$ state, the result reads
\begin{equation}
   |\psi\rangle_{LR \rightarrow RL + LR} = \frac{1}{\sqrt{t^{4} + u^{4}}}\left( t^{2}|RL\rangle + u^{2}|LR\rangle\right)
   .
\end{equation}

A relevant feature in the above result is the cancellation of color degrees of freedom for all non-zero amplitudes. To be precise, for those color amplitudes which are non-vanishing, the balance between $LR$ and $RL$ states is not affected by gauge indices.
In other words, the final state generated is the same independently of the color of the gluons involved in the interaction, which implies that the color degrees of freedom are neutral witnesses for any quantum information quantity computed from the final state wavefunction, including the entanglement measured with the concurrence or the violation of a Bell inequality.
This is in no contradiction with the fact that output colors have different probabilities, as shown in Eq.~\eqref{eq:MRL} and dictated by the color structure functions.

The simple form for the scattering of polarizations is the result of cancellations between the $t$ and $u$ channels {\sl vs.} the quartic vertex contribution. The $s$-channel does not generate entanglement. As a matter of fact, it is never necessary to use the Jacobi identity for the structure constants. The simple form  of the final result emerges for any $SU(N)$ group.

\section{Generation of entanglement}\label{sec:generation_entanglement}

The generation of entanglement in gluon scattering can now be quantified using the concurrence, defined in Eq.~\eqref{eq:delta}. For the final state showed in Eq.~\eqref{eq:finalRL}, we obtain 
\begin{equation}
    \Delta_{RL \rightarrow RL + LR} =   \frac{2 t^{2}u^{2}}{t^{4} + u^{4}}
    \label{eq:cocurrenceMandelstam}
\end{equation}
which, in the center of mass frame, corresponds to 
\begin{equation}
    \Delta_{RL \rightarrow RL + LR}  =   \frac{2 \tan^{4}\left(\frac{\theta}{2}\right)}{1 + \tan^{8}\left(\frac{\theta}{2}\right)}, 
\end{equation}
where $\theta$ is the COM angle.

An identical result is obtained starting from an $LR$ state.
Therefore, concurrence for the polarizations of a process mediated by the strong force only depends on the scattering angle.  

A first observation about the above result is that concurrence for gluon polarization found in Eq. \eqref{eq:cocurrenceMandelstam} takes the exact same form as the one for helicities in identical fermionic scattering computed in Ref. \cite{CerveraLierta2017maximal}. There, the contributions from $t$ and $u$ channels are indistinguishable, bringing the possibility of maximal entanglement for any mass of the fermions. In the case of gluon dynamics, the variable $s$ is not contributing at all to the amplitude, and the four-vertex channel cancels some piece of the $u$ and $t$ channels. Thus, although the result is the same for indistinguishable fermions and for distinguishable colored gluons, the underlying mechanisms to achieve maximal entanglement are slightly different.

To obtain a maximal entangled state, concurrence needs to be $\Delta = 1$. From Eq.\eqref{eq:cocurrenceMandelstam} it follows that this happens only when $\theta = \frac{\pi}{2}$, i.e. $t = u$. In this scenario, for every initial two-gluon product state with opposite polarizations, the final state will be always maximally entangled, no matter the color charges of the initial state if not identical. 
Then, the maximal entangled final states take the form
\begin{equation}
   |\psi^{+}\rangle = \frac{1}{\sqrt{2}}\left(|RL\rangle + |LR\rangle\right).
   \label{eq:bell}
\end{equation}
in both cases. The emergence of the zero component of a triplet is natural due to the fact that gluons are bosons. In fermionic scattering, the singlet is obtained, showing again the different nature of both processes.

\section{Exploring a MaxEnt principle}\label{sec:MaxEntk}

The results obtained for the entanglement in gluon scattering show that the detailed mechanism to achieve maximal entanglement is deeply rooted in the interplay between $t$, $u$ and quartic-vertex channels. It is natural to explore departures from this fine balance. A more ambitious point of view can be stated in the form of a principle: The laws of Nature must be able to generate maximal entanglement in scattering processes of incoming particles which are not entangled. This is tantamount to say that Nature must be exposed to Bell inequalities, that should be violated. This is to say that Nature should not be describable by a classical theory.
Following Ref. \cite{CerveraLierta2017maximal}, we refer to this idea as a MaxEnt Principle, that may constrain the structure of interactions.

To investigate this idea, we modify the balance between the 3- and 4-gluon interactions. This is done applying a weight $k$ to the 4-gluon vertex, that leads to a total amplitude
\begin{equation}
    \mathcal{M} = \mathcal{M}_{s} + \mathcal{M}_{t} + \mathcal{M}_{u} + k \mathcal{M}_{4}.
    \label{eq:addingk}
\end{equation}
As we shall discuss shortly, the outcomes for $k \not= 1$ correspond to interactions that are not gauge invariant. 
To be precise, we break gauge invariance in the interaction term of the QCD lagrangian only. Therefore, other Feynman rules such as the gluon propagators, or the gluons degrees of freedom (they are massless bosons) are not affected by this modification. Although there are other ways to break gauge invariance, we chose this one as we consider it a minimal gauge symmetry braking that allow us to explore the power of imposing MaxEnt in a more general theory.
The values for each amplitude as a function of $k$ are listed in App. \ref{app:amplitudes4k}. 

By repeating the computation in the previous section, we now find that the only value of $k$ for which the generation of entanglement is independent of the color and for all values of $\theta$, is the $SU(N)$ gauge invariant case $k=1$. 

\begin{figure}[t!]
    \centering
    \includegraphics[width=0.5\linewidth]{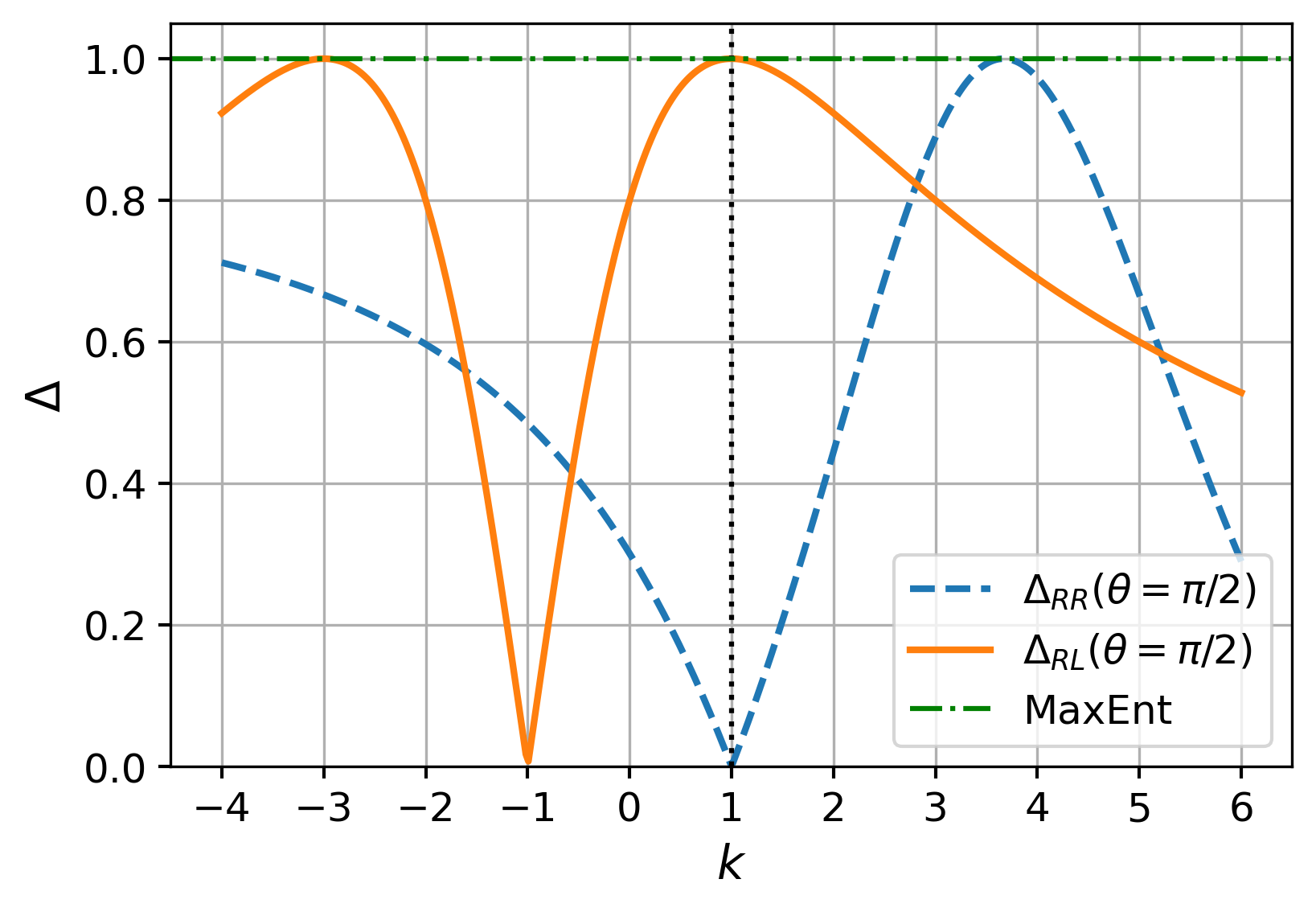}
    \caption{Concurrence as a function of the 4 vertex parameter $k$ and COM angle $\theta=\pi/2$. Maximal entanglement is achieved for the QCD solution $k=1$ and initial gluon polarization of $|RL\rangle$, but other unphysical solutions are also obtained for $k=-3$ and also for $k=11/3$ for an initial polarization of $|RR\rangle$. Equivalent results are obtained for initial polarizations $|LR\rangle$ and $|LL\rangle$.}
    \label{fig:MaxEnt}
\end{figure}

Let us now concentrate in the case we fix the scattering angle to $\theta = \pi/2$, the concurrence for any initial polarization becomes independent of the color degree of freedom for any value of $k$. In this scenario, the concurrence for initial states of opposite polarizations read
\begin{equation}
    \Delta_{RL \rightarrow RL + LR}  =  \left|\frac{8 (k +1)}{5 + 2k + k^{2}}\right| .
\end{equation}
The computation shows that only the value $k=1$ leads to a final maximal entangled state, i.e. $\Delta=1$, which corresponds to the theory respecting gauge symmetry. This solution is an isolated point as shown in Fig.~\ref{fig:MaxEnt}, and also suppresses the $LL\to LL+RR$ process, since
\begin{equation}
    \Delta_{RR \rightarrow LL + RR}= \Delta_{LL \rightarrow LL + RR}   = 2 \left| \frac{2 (k -1)(k-7)}{93-34k+5k^{2}}\right| .
\end{equation}
These results show how fine-tuned is the gauge invariant Lagrangian for gluon dynamics in terms of how much entanglement can be created. There are no flat directions. The gauge invariant theory appears as an isolated point of maximum entanglement with respect to small variations of the parameter $k$.

It is possible to analyze other  scenarios which are departures of the standard theory. For an initial $|RL\rangle$ polarization,
there is a second solution for $k = -3$. Let's write explicitly the final state as a function of $k$ for an initial $|RL\rangle$ polarization and at $\theta=\pi/2$: 
\begin{equation}
    |\psi\rangle = \frac{1}{\sqrt{(k-1)^2+(k+3)^2}}\left((k-1)|\phi^{+}\rangle + (k+3)|\psi^{+}\rangle\right), 
\end{equation}
where $|\phi^{+}\rangle=(|RR\rangle +|LL\rangle)/\sqrt{2}$. The final state oscillates between two maximally entangled states: one that corresponds to an unphysical scenario (as the theory would not obey Ward identities that preserve the correct degrees of freedom at higher orders of perturbation theory), $|\phi^{+}\rangle$, and the QCD solution from Eq.~\eqref{eq:bell}.

It is also possible to check whether the initial state $RR$ can generate entanglement in a non gauge invariant theory. In that case, maximal entanglement would be attained at $k=11/3$. The final states would be
\begin{equation}
    |\psi\rangle = \frac{1}{2 \sqrt{5}}\Big(|RL\rangle + |LR\rangle + 3 (|RR\rangle - |LL\rangle)\Big).
\end{equation}
A simple rotation of one of the polarizations can show that this state can be transformed into one of the Bell states.

In summary, although a MaxEnt principle in gluon dynamics at tree level is not enough to completely restrict the gluon interaction to the gauge invariant case, it does single out $k=1$ as an isolated point where maximal entanglement is achieved, and gauge symmetry is recovered.

\section{Conclusions}\label{sec:conclusions}

Fundamental interactions generate entangled states by means of  indistinguishability of the relevant degrees of freedom involved. 

In the case of QED, the superposition of the $t$ and $u$-channel is at the core of the generation of entanglement for indistinguishable fermions. In the case particle-antiparticle collisions, entanglement emerges through the $s$-channel, where the virtual photon couples identically to the two options for helicities of the outgoing particles. 

Gluon dynamics poses a different problem, as in-going, out-going and virtual intermediate particles are bosons with a color index on top of the polarization. The net effect on the entanglement of polarization degrees of freedom requires to add in superposition the contribution of all $s$, $t$, $u$ and quartic-vertex channels. A detailed computation shows that entanglement among polarizations of the gluons is only generated when the initial product state presents opposite polarizations. It also shows that maximal entanglement is obtained when outgoing particles are in the transverse plane. 

A non-obvious result coming from this computation is that the amount of entanglement produced in gluon collisions does not depend on the color charge of the gluons. For all combinations of initial and final color indices which are allowed, only the scattering angle of the final state matters, and maximal entangled states arise when $t = u$, that is  when the final gluons trajectories are perpendicular to the initial ones. 

The generation of maximal entanglement shows that nature is quantum and e.g. QCD
cannot be reproduced by a classical theory based on local determinism. In other words,
if violation of would-be Bell inequalities is mandatory, then gluon dynamics is only describable
by a quantum theory.
The production of entanglement in gluon scattering is independent of the gauge group. However, a small departure of the gauge-tuned relation between the three- and four-gluon vertices would entail a reduction of entanglement.
We analyze this possibility by breaking gauge invariance in the interaction term by modifying the balance between the 3- and 4-gluon vertex. Other ways of exploring this gauge symmetry emergence from MaxEnt can be explored, but what we can observe is that those gauge symmetry breaking choices related with the color degrees of freedom will be blind to such modifications.
Therefore, even if a possible Principle of MaxEnt does not select a particular gauge group as preferred by Nature, Nature fulfills such a Principle so that universality of entanglement on gauge theories emerges.

\section{Acknowledgements}
C. N.  thanks the Technology Innovation Institute for granting an internship in the Quantum Research Center. A.C.-L. acknowledges funding from Grant RYC2022-037769-I funded by MICIU/AEI/10.13039/501100011033 and by “ESF+".

\clearpage{}

\appendix
\section{Conventions}\label{app:conventions}

In this section we list all the convention used in this work. We start by stating the kinematics of the process and then define the Feynman rules used to compute the scattering amplitudes.

\subsection{ Kinematics}
We work in the center of mass (CM) frame, using natural units $c = \hbar = 1$ and the metric signature $\eta^{\mu\nu} = \textrm{diag}(+---)$. We consider the process to take place in the xz-plane, with the momentum of the incoming particles along the z-axis. Then, the momenta are 
\begin{equation}
\begin{aligned}
    & p_{1}^{\mu} = (p,0,0,p) \\
   & p_{2}^{\mu} = (p,0,0,-p),
   \label{inp}
\end{aligned}
\end{equation}
for the incoming gluons and 
\begin{equation}
\begin{aligned}
    & p_{3}^{\mu} = (p,p \sin\theta,0,p \cos\theta) \\
    & p_{4}^{\mu} = (p,-p \sin\theta,0,-p \cos\theta),
    \label{outp}
\end{aligned}
\end{equation}
for the outgoing ones, where $\theta$ is the scattering angle. 

The circular polarization vectors for a gluon having a momentum $k^{\mu}$ are defined as 
\begin{equation*}
    \epsilon^{\mu}(k^{\mu},\lambda)= (0,\vec{\epsilon}),
\end{equation*}
where
\begin{equation*}
    \vec{\epsilon} = \frac{-\lambda}{\sqrt{2}}\left(\cos\theta
    \cos\phi -i\lambda \sin\phi, \cos\theta \sin\phi + i\lambda
    \cos\phi, -\sin\theta\right)
\end{equation*}
and $\lambda = \pm 1$, that correspond to R and L, respectively. Then, the polarization vectors for the gluons having momenta Eq. (\ref{inp}) and Eq. (\ref{outp}) take the form
\begin{equation}
\begin{aligned}
    &\epsilon^{\mu} (p_{1}) = -\frac{\lambda_{1}}{\sqrt{2}} \left(0, 1, i\lambda_{1}, 0\right) \\
    &\epsilon^{\mu} (p_{2}) = -\frac{\lambda_{2}}{\sqrt{2}}\left(0, 1, -i\lambda_{2}, 0\right) \\
    & \epsilon^{\mu} (p_{3}) = -\frac{\lambda_{3}}{\sqrt{2}}\left(0, \cos \theta, -i\lambda_{3}, - \sin\theta\right) \\
    & \epsilon^{\mu} (p_{4}) = -\frac{\lambda_{4}}{\sqrt{2}}\left(0, \cos \theta, i\lambda_{4}, - \sin\theta\right).
\end{aligned}
\end{equation}
Each polarization vector is transverse to the corresponding gluon momentum,  $\epsilon^{\mu}(k) k_{\mu} = 0$, due to the massless nature of gluons.

Finally, we define the Mandelstam variables as
\begin{equation}
    \begin{aligned}
        & s = (p_{1} + p_{2})^{2} = (p_{3} + p_{4})^{2} \\
        & t = (p_{1} - p_{3})^{2} = (p_{2} - p_{4})^{2}  \\
        & u = (p_{1} - p_{4})^{2} = (p_{2} - p_{3})^{2} ,
        \label{mandelstam}
    \end{aligned}
\end{equation}
where $s$ is the squared center-of-mass energy and $t$ and $u$ are the squared four-momentum transfer.

\subsection{Feynman rules}

Feynman rules are mathematical expressions that represent terms in the Lagrangian of the theory at work: free external particles and the possible interactions between them. In this section, we list the Feynman rules of the gluons self-interactions.

\vspace{0.3cm}

\textit{ Three gluon vertex}
\\
\begin{tikzpicture}
  \begin{feynman}
    \vertex (a) {\(g_{\rho}^{c}\)};
    \vertex [below=of a] (b);
    \vertex [below left=of b] (f1) {\(g_{\mu}^{a}\)};
    \vertex [below right=of b] (f2) {\(g_{\nu}^{b}\)};

    \diagram* {
      (a) -- [gluon, momentum'={$k$}] (b) -- [gluon,  rmomentum'={$p_{1}$}] (f1),
      (f2) -- [gluon,  momentum'={$p_{2}$}] (b),
    };
  \end{feynman}
\end{tikzpicture}
 \begin{equation*}
   =  g f^{abc} [\eta^{\mu\nu}\left(p_{1}-p_{2}\right)^{\rho} 
    + \eta^{\nu\rho}\left(p_{2}-p_{3}\right)^{\mu} 
   +\eta^{\rho\mu}\left(p_{3}-
p_{1}\right)^{\nu}]
 \end{equation*}

\textit{ Four gluon vertex}

\begin{tikzpicture}
  \begin{feynman}
    \vertex (a) {\(g_{\mu}^{a}\)};
    \vertex [below=of a] (b);
    \vertex [left=of b] (f1) {\(g_{\kappa}^{a'}\)};
    \vertex [right=of b] (f2) {\(g_{\sigma}^{b'}\)};
     \vertex [below=of b] (f3) {\(g_{\nu}^{b}\)};

    \diagram* {
      (a) -- [gluon, momentum'={$p_{1}$}] (b),
      (f1) -- [gluon,  momentum'={$p_{3}$}] (b),
      (f2) -- [gluon,  momentum'={$p_{4}$}] (b),
      (f3) -- [gluon,  momentum'={$p_{2}$}] (b),
    };
  \end{feynman}
\end{tikzpicture}
\begin{equation*}
         =  -i g^{2} [f^{abc}f^{a'b'c}(\eta^{\mu\kappa}\eta^{\nu\sigma}-\eta^{\mu\sigma}\eta^{\nu\kappa}) 
         + f^{aa'c}f^{bb'c}(\eta^{\mu\nu}\eta^{\kappa\sigma}-\eta^{\mu\sigma}\eta^{\nu\kappa}) 
         + f^{ab'c}f^{ba'c}(\eta^{\mu\nu}\eta^{\kappa\sigma}-\eta^{\mu\kappa}\eta^{\nu\sigma})]
\end{equation*}

\section{Polarized amplitudes \texorpdfstring{$gg \rightarrow gg$}{gg to gg}}\label{app:amplitudes}

Using the Feynman rules in App. \ref{app:conventions} we obtain the amplitudes for each channel. In the following section we evaluate these amplitudes for the four channels involved in the gluon scattering process, and obtain the results for all combinations of initial and final polarizations. In all of them, $g$ is the strong coupling constant, $f^{ijk}$ the structure constants and $s$, $t$, $u$ the Mandelstam variables.

\subsection{\texorpdfstring{$s$-channel}{s-channel}}

The scattering amplitude of the $s$-channel takes the form
\begin{equation}
\begin{aligned}
        i\mathcal{M}_{s} = &  -i \frac{g^{2}f^{abc}f^{a'b'c}}{s}\epsilon_{\mu}(p_{1})\epsilon_{\nu}(p_{2})\epsilon_{\kappa}^{*}(p_{3})\epsilon_{\sigma}^{*}(p_{4}) \\
       & \left[\eta^{\mu\nu}(p_{1}-p_{2})^{\rho} + 2\left(\eta^{\nu\rho}p_{2}^{\mu} - \eta^{\rho\mu}p_{1}^{\nu}\right) \right] \\
       & \left[\eta^{\kappa\sigma}(p_{4}-p_{3})_{\rho} + 2\right( \eta_{\rho}^{\kappa}p_{3}^{\sigma}-\eta^{\sigma}_{\rho}p_{4}^{\kappa}\left) \right].
\end{aligned}
\end{equation}
The four non-zero amplitudes correspond to the case where both gluons in the initial product state share the same polarization, and give rise to a final state where, likewise, the polarization is the same for both gluons. Specifically, these processes correspond to $ RR\rightarrow RR$, $RR\rightarrow LL$, $LL \rightarrow RR$ and $LL \rightarrow LL$, all of which share the same amplitude value,

\begin{equation}
    \mathcal{M}_{s} = g^{2}f^{abc}f^{a'b'c} \left(\frac{u-t}{s}\right).
\end{equation}

\subsection{\texorpdfstring{$t$-channel}{t-channel}}

The scattering amplitude of the $t$-channel takes the form
\begin{equation}
\begin{aligned}
        i\mathcal{M}_{t} = & -i \frac{g^{2}f^{aa'c}f^{bb'c}}{t}\epsilon_{\mu}(p_{1})\epsilon_{\nu}(p_{2})\epsilon_{\kappa}^{*}(p_{3})\epsilon_{\sigma}^{*}(p_{4}) \\
       & \left[\eta^{\mu\kappa}(p_{1}+p_{3})^{\rho} -2\right(\eta^{\kappa\rho}p_{3}^{\mu} + \eta^{\rho\mu}p_{1}^{\kappa}\left) \right] \\
       & \left[\eta^{\nu\sigma}(p_{2}+p_{4})_{\rho} - 2\left(\eta^{\sigma}_{\rho}p_{4}^{\nu} + \eta_{\rho}^{\nu}p_{2}^{\sigma} \right) \right].
\end{aligned}
\end{equation}
In this case none of them equals zero, but some processes give rise to the same value. We obtain four different values
\begin{eqnarray}
    \mathcal{M}_{RR \rightarrow RR} &=& \mathcal{M}_{LL \rightarrow LL} = 
    -g^2 F_{1}\left(2\frac{4t+u}{s}+\frac{tu}{s^2}\right)\frac{u}{t}, \nonumber\\
    \mathcal{M}_{RL \rightarrow RL} &=& \mathcal{M}_{LR\rightarrow LR} = 
    g^2 F_{1}\frac{t+2u}{s}\frac{u^{2}}{ts}, \nonumber\\
    \mathcal{M}_{RR \rightarrow LL} &=& \mathcal{M}_{LL \rightarrow RR} = \mathcal{M}_{RL \rightarrow LR} = 
    \mathcal{M}_{LR\rightarrow RL} = 
    g^2 F_{1} \frac{t+2u}{s}\frac{t}{s}, \nonumber \\
    \mathcal{M}_{ \substack{RR \\ LL}\rightarrow  {\substack{RL \\ 
        LR}}} &=& \mathcal{M}_{ \substack{RL \\
        LR} \rightarrow  \substack{RR \\ LL}} = 
          - g^{2}F_{1} \frac{tu}{s^{2}},
\end{eqnarray}
where $F_{1}\equiv f^{aa'c}f^{bb'c}$ and we use $s+t+u=0$.

\subsection{\texorpdfstring{$u$-channel}{u-channel}}

The scattering amplitude of the $u$-channel takes the form
\begin{equation}
\begin{aligned}
        i\mathcal{M}_{u} = &  -i \frac{g^{2}f^{ab'c}f^{ba'c}}{u}\epsilon_{\mu}(p_{1})\epsilon_{\nu}(p_{2})\epsilon_{\kappa}^{*}(p_{3})\epsilon_{\sigma}^{*}(p_{4}) \\
       & \left[\eta^{\mu\sigma}(p_{1}+p_{4})^{\rho} -2 \left( \eta^{\sigma\rho}p_{4}^{\mu} + \eta^{\rho\mu}p_{1}^{\sigma} \right)\right] \\
       & \left[\eta^{\nu\kappa}(p_{2}+p_{3})_{\rho} - 2\left(\eta^{\kappa}_{\rho}p_{3}^{\nu} + \eta_{\rho}^{\nu}2p_{2}^{\kappa} \right) \right].
\end{aligned}
\end{equation}
For this channel we also obtain that none of the amplitudes equals zero and that some processes give rise to the same value. As in the t-channel, we obtain four different values
\begin{eqnarray}
    \mathcal{M}_{RR \rightarrow RR} &=& \mathcal{M}_{LL\rightarrow LL} = 
    -g^2F_{2}\left(2\frac{4u+t}{s}+\frac{tu}{s^2}\right)\frac{t}{u}, \nonumber\\
    \mathcal{M}_{RL \rightarrow LR} &=& \mathcal{M}_{LR \rightarrow RL} = 
    g^2F_{2}\frac{2t+u}{s}\frac{t^{2}}{us}, \nonumber\\
     \mathcal{M}_{RR\rightarrow LL} &=& \mathcal{M}_{LL \rightarrow RR} = \mathcal{M}_{RL \rightarrow RL} = \mathcal{M}_{LR\rightarrow LR} = 
         g^2F_{2}\frac{2t+u}{s}\frac{u}{s}, \nonumber\\
    \mathcal{M}_{\substack{RR \\ LL}\rightarrow \substack{RL \\ LR}} &=& \mathcal{M}_{\substack{RL \\
        LR} \rightarrow \substack{RR \\ LL}} =  - g^{2}F_{2} \frac{tu}{s^{2}},
\end{eqnarray}
where $F_{2}\equiv f^{ab'c}f^{ba'c}$.

\subsection{4-point}
Lastly, the scattering amplitude of the 4-point channel takes the form
\begin{equation}
\begin{aligned}
        i\mathcal{M}_{4} = & -i g^{2}[f^{abc}f^{a'b'c}\left[(\epsilon_{1}\cdot\epsilon_{3}^{*})(\epsilon_{2}\cdot\epsilon_{4}^{*})-(\epsilon_{1}\cdot\epsilon_{4}^{*})(\epsilon_{2}\cdot\epsilon_{3}^{*})\right] \\
       & + f^{aa'c}f^{bb'c}\left[(\epsilon_{1}\cdot\epsilon_{2})(\epsilon_{3}^{*}\cdot\epsilon_{4}^{*})-(\epsilon_{1}\cdot\epsilon_{4}^{*})(\epsilon_{2}\cdot\epsilon_{3}^{*})\right] \\
       & + f^{ab'c}f^{ba'c}\left[(\epsilon_{1}\cdot\epsilon_{2})(\epsilon_{3}^{*}\cdot\epsilon_{4}^{*})-(\epsilon_{1}\cdot\epsilon_{3}^{*})(\epsilon_{2}\cdot\epsilon_{4}^{*})\right]].        
\end{aligned}
\end{equation}
For this channel we obtain five different non-zero values, that are
\begin{eqnarray}
    \mathcal{M}_{RR \rightarrow RR} &=& \mathcal{M}_{LL \rightarrow LL} = 
        g^2\left(F_{3}\frac{u-t}{s}-F_{1}\frac{2t+u}{s}\frac{u}{s} -F_{2}\frac{2u+t}{s}\frac{t}{s}\right), \nonumber\\
        \mathcal{M}_{RR \rightarrow LL} &=& \mathcal{M}_{LL\rightarrow RR} = 
       - g^2\left(F_{3}\frac{u-t}{s}+F_{1}\frac{2u+t}{s}\frac{t}{s} +F_{2}\frac{2t+u}{s}\frac{u}{s}\right), \nonumber\\
        \mathcal{M}_{RL \rightarrow RL} &=& \mathcal{M}_{LR \rightarrow LR} = g^2\left(F_{1}+F_{2}\right)\left(\frac{u}{s}\right)^2, \nonumber\\
        \mathcal{M}_{RL \rightarrow LR} &=& \mathcal{M}_{LR \rightarrow RL} = 
        g^2 \left(F_{1}+F_{2}\right)\left(\frac{t}{s}\right)^2, \nonumber\\
    \mathcal{M}_{\substack{RR \\ LL}\rightarrow \substack{RL \\ LR}} &=& \mathcal{M}_{\substack{RL \\
        LR} \rightarrow \substack{RR \\ LL}} =         g^2\left(F_{1}+F_{2}\right)\frac{tu}{s^2},
\end{eqnarray}
where $F_{3}\equiv f^{abc}f^{a'b'c}$.

\section{Total amplitudes with vertex modification}\label{app:amplitudes4k}

In this section we list the total scattering amplitudes obtained when adding a weight $k$
to the 4-gluon vertex, Eq.~\eqref{eq:addingk}. 

We start with the amplitudes that had non-zero value for $k=1$. We obtain
\begin{eqnarray}
    \mathcal{M}_{RR \rightarrow RR} &=& \mathcal{M}_{LL \rightarrow LL} =  \mathcal{M}_{k=1} 
         +\frac{g^{2}}{2} (k-1) \left( 2 F_{3} \left(\frac{u-t}{s}\right) 
         -F_{1}\frac{2u(2t+u)}{s^2} - F_{2}\frac{2t(2u+t)}{s^2}\right), \nonumber\\
        \mathcal{M}_{RL \rightarrow RL} &=& 
        \mathcal{M}_{LR \rightarrow LR} =  \mathcal{M}_{k=1}
         + g^{2} (k-1) \left(F_{1} + F_{2}\right) \left(\frac{u}{s}\right)^{2}, \nonumber\\
        \mathcal{M}_{RL\rightarrow LR} &=& 
        \mathcal{M}_{LR \rightarrow RL} =  \mathcal{M}_{k=1} 
         + g^{2} (k-1) \left(F_{1} + F_{2}\right) \left(\frac{t}{s}\right)^{2},
\end{eqnarray}
where $\mathcal{M}_{k=1}$ are the amplitudes in Eq.~\eqref{eq:MRR} and Eq.~\eqref{eq:MRL} and $F_{1}\equiv f^{aa'c}f^{bb'c}$, $F_{2}\equiv f^{ab'c}f^{ba'c}$ and $F_{3}\equiv f^{abc}f^{a'b'c}$.

The remaining amplitudes are
\begin{eqnarray}
    \mathcal{M}_{RR\rightarrow LL} &=& \mathcal{M}_{LL \rightarrow RR} = 
         - \frac{g^{2}}{2} (k-1) \left( 2 F_{3} \left(\frac{u-t}{s}\right) 
         + F_{1} \frac{2t(2u+t)}{s^2}
         + F_{2}\frac{2u(2t+u)}{s^2} \right), \nonumber\\
    \mathcal{M}_{ \substack{RR \\ LL}\rightarrow  \substack{RL \\
        LR}} &=&  \mathcal{M}_{\substack{RL \\ LR} \rightarrow  \substack{RR \\ LL}} = 
          g^{2} (k-1) \left(F_{1} + F_{2} \right)\frac{tu}{s^{2}}.
\end{eqnarray}

\bibliographystyle{JHEP}
\bibliography{biblio.bib}

\end{document}